\documentclass[12pt,preprint]{aastex}
\newcommand{\ie}{i.e.,}
\newcommand{\eg}{e.g.,}

\newcommand{\etal}{et~al.\ }
\newcommand{\ltsima}{$\; \buildrel < \over \sim \;$}
\newcommand{\simlt}{\lower.5ex\hbox{\ltsima}}
\newcommand{\gtsima}{$\; \buildrel > \over \sim \;$}
\newcommand{\simgt}{\lower.5ex\hbox{\gtsima}}
\newcommand{\kms}{km s$^{-1}$}

\def\parcmin{{\tt '}\mskip -6.0mu.\,}
\newcommand{\oiii}{[\ion{O}{3}]~$\lambda$5007}
\def\rquart{$r^{1/4}$}
\def\parcmin{{\tt '}\mskip -6.0mu.\,}

\begin{document}
 
\title{Intracluster Planetary Nebulae in the Virgo Cluster 
III: Luminosity of the Intracluster Light 
and Tests of the Spatial Distribution}

\author{John J. Feldmeier\altaffilmark{1,2,3}}
\affil{Department of Astronomy and Astrophysics, Penn State University,
525 Davey Lab, University Park, PA 16802}
\email{johnf@bottom.astr.cwru.edu}

\author{Robin Ciardullo\altaffilmark{1}}
\affil{Department of Astronomy and Astrophysics, Penn State University,
525 Davey Lab, University Park, PA 16802}
\email{rbc@astro.psu.edu}

\author{George H. Jacoby}
\affil{WIYN Observatory
\\ P.O. Box 26732, Tucson, AZ 85726}
\email{gjacoby@wiyn.org}

\and

\author{Patrick R. Durrell}
\affil{Department of Astronomy and Astrophysics, Penn State University,
525 Davey Lab, University Park, PA 16802}
\email{pdurrell@astro.psu.edu}

\altaffiltext{1}{Visiting Astronomer, Kitt Peak National Observatory, 
National Optical Astronomy Observatory, which is operated by 
the Association of Universities for Research in Astronomy, Inc. 
(AURA) under cooperative agreement with the National Science Foundation.}

\altaffiltext{2}{present address: Department of Astronomy, Case 
Western Reserve University, 10900 Euclid Ave, Cleveland, OH 44106}

\altaffiltext{3}{NSF Astronomy and Astrophysics Postdoctoral Fellow}

\begin{abstract}

Intracluster planetary nebulae are a useful tracer of the evolution 
of galaxies and galaxy clusters.
We analyze our catalog of 318 intracluster planetary nebulae candidates
found in 0.89 square degrees of the Virgo cluster.  
We give additional evidence for the great depth of the Virgo cluster's
intracluster stellar population, which implies that the bulk
of the intracluster stars come from late-type galaxies and dwarfs.  We also
provide evidence that the intracluster stars are clustered on the
sky on arcminute scales, in agreement with tidal-stripping 
scenarios of intracluster star 
production.  Although significant systematic uncertainties 
exist, we find that the average fraction of intracluster starlight 
in the Virgo is $15.8\% \pm 3.0 \mbox{(statistical)} \pm 5.0 
\mbox{(systematic)}$, and may be 
higher if the intracluster stars have a large 
spatial line-of-sight depth.  We find that the intracluster 
star density changes little with radius or projected density over the
range surveyed.  These results, along with other intracluster star
observations, imply that intracluster star 
production in Virgo is ongoing and consistent with the cluster's 
known dynamical youth.  
 
\end{abstract}
 

\section{Introduction}

The concept of intracluster starlight was first proposed by \citet{zwicky1951},
who claimed to detect excess light between the galaxies of the Coma
cluster.  Follow-up photographic searches for intracluster luminosity in
Coma and other rich clusters \citep[e.g.,][see V\'ichez-G\'omez 1999 and
Feldmeier 2000 for reviews]{wel1971,mel1977} produced mixed 
results, and it was not until the advent 
of CCDs that more precise estimates of the amount of intracluster starlight 
were made \citep[e.g.,][]{uson1991b,vg1994,bern1995,gon2000}. 
However, even these modern measurements carry a substantial uncertainty, due
to the extremely low surface brightness of the phenomenon.  Typically
the brightest intracluster 
regions are less than $\sim 1\%$ of the brightness of the
night sky in the $V$ band \citep[e.g.,][]{icl1}, 
and measurements of this luminosity must also 
contend with the problems presented by scattered light from 
nearby bright objects and the contribution of faint discrete sources.

Recently, a great deal of progress has been made in the study of 
intracluster light (ICL).  Our ability to measure the ICL in Abell clusters has
grown to the point that reliable surface photometry is now possible at
levels several magnitudes (0.1\% to .01\% in the $V$ and $I$ bands) 
lower than that of the sky \citep[e.g.,][]{tyson1998,gon2000,icl1}.
These studies are making it possible to study the evolution of the ICL as a 
function of redshift, cluster substructure, galactic density, and
galactic morphology.  Meanwhile, high-resolution N-body models now
have the ability to follow hundreds of interacting cluster galaxies within
a cosmological context \citep{harass,dub1998,nap2003,mihos2004b,willman2004}.  
These high-quality simulations 
are making testable predictions concerning the production of intracluster
light. When combined with earlier and recent semi-analytic studies
\citep{gal1972,mer1983,rich1983,mil1983,mer1984,muccione2004} 
there is now sufficient theoretical insight to begin to interpret 
the observations. 

Another major advance in the study of the ICL is our growing ability to
study individual intracluster objects such as globular clusters 
\citep{igc1995,igc2003} and in particular,  
individual intracluster stars, specifically planetary
nebulae (IPNe), red giants (IRGs; \citet{ftv1998,durr2002}) 
and supernovae (ISNe; 
\citet{galyam2003}).  Perhaps the
most useful of these objects are the planetaries.  
IPNe have several unique features that make them ideal for 
probing intracluster starlight 
\citep[for a complete review, see][]{ipnreview}.  
In the light of \oiii, planetary nebulae (PNe) are
extremely luminous, and can be observed out to distances of $\sim 25$ Mpc
with present day telescopes.  Planetary nebulae are also excellent 
tracers of stellar luminosity (Ciardullo, Jacoby, \& Ford 1989; 
Ciardullo \etal 1989), and their number counts can be scaled
to produce estimates of the total amount of unseen light. 
In addition, through 
the \oiii~planetary nebula luminosity function (PNLF), PNe are 
precise distance indicators (Jacoby \etal 1992; Ciardullo 2002b); 
therefore, the observed shape of the PNLF provides information on the 
line-of-sight distribution of the intracluster stars.  
Finally, since the \oiii~emission is in a narrow line, 
IPN velocities can 
be determined via moderate ($\lambda /\Delta \lambda \sim 2000$) resolution 
spectroscopy, making kinematical studies of the ICL possible.  

In this paper, we utilize the 318 IPN candidates detected in
our photometric survey of Virgo (Feldmeier, Ciardullo, \& Jacoby 1998; 
hereafter Paper~I; Feldmeier \etal 2003a; hereafter Paper~II) to 
place limits on the amount and distribution of intracluster starlight 
in that cluster.  In \S 2, we review the observations and reduction procedures
which produced the IPN catalog.  In \S 3, we use the IPN data to estimate
the three-dimensional structure of Virgo's diffuse stellar component.  
We compare the distribution of ICL to that of the galaxies and the hot 
intracluster medium, and present evidence that Virgo's IPNe are clumped
on small scales.  In \S 4, we compute the total amount of intracluster
starlight that must be present in Virgo, discuss the systematic
uncertainties present in this number, and compare the large scale
distribution of IPNe with that of two common parameters of 
cluster dynamical evolution: cluster radius and projected galaxy density.  
We discuss the implications of our results in \S 5 and combine our data
with the observation of intracluster red giants, to place
constraints on the population of the intracluster stars and on their
origins.  We stress that not all of the 318 candidates are 
genuine IPNe: we estimate (Ciardullo \etal 2002a) 
that $\sim$ 20\% of the IPN candidates 
are likely background objects that are themselves of considerable 
scientific interest (\eg~Kudritzki \etal 2000; Rhoads \etal 2000; 
Stern \etal 2000).  

For this paper, we adopt $\sim 15$~Mpc as the distance to the Virgo cluster
core.  This number is consistent with the PNLF distances to the elliptical
galaxies of the core \citep{pnlf5, m87ipn}, and enables us to compare our 
IPN measurements directly with the PNLF observations of ellipticals.  We note
that such a distance is $\sim 0.7$~Mpc smaller than the mean distance 
estimated from Cepheid variables by the {\sl HST\/} 
Distance Scale Key Project and others 
(Pierce \etal 1994; Saha \etal 1997; Freedman \etal 2001).  
Such a discrepancy is not serious, especially since there is growing evidence
that the ellipticals of Virgo are not necessarily at the same distance
as the spirals \citep{wb2000, dgc2001}.  Additionally, the median distances
of the two methods are consistent within the errors, 
and the difference in the mean is solely due to the most distant 
galaxy NGC~4639 (D = $20.8^{+2.8}_{-2.5}$; Freedman \etal 2001).  
In any case, all quantities in this paper that are dependent on this 
distance will be expressed as a function of $D_{15}$, 
where $D_{15} = D_{{\rm Virgo Core}} / 15$~Mpc.

\section{Observations}

The detection, photometry, and astrometry of the 318 IPN candidates were 
reported in full in Paper~II.  Here, we briefly summarize the results.  
The images for this survey were taken over the course of three
observing runs from March 1997 to March 1999 at the Kitt Peak National
Observatory (KPNO) using the 4~m telescope and the Prime Focus CCD camera 
for runs 1 and 2 (PFCCD) and the NOAO MOSAIC~I imager 
(MOSAIC; Muller \etal 1998) for run 3.  The data are composed
of images taken through an \oiii~narrow-band filter, and a medium-band
continuum filter.  IPN candidates appear as point sources in the former
filter, but disappear in the latter filter.  The IPN candidates were detected
semi-automatically using two separate methods, and objects in, or
adjacent to, Virgo cluster galaxies have been culled from our sample.  
In addition, any source known to be present in the continuum,
or shown to be a background galaxy through follow-up spectroscopy 
has been removed.  We
stress, however, that only 10\% of the sources have follow-up 
spectra, and those spectra that
do exist have low signal-to-noise.  We will return to this point later.

\begin{figure}
\figurenum{1}
\epsscale{1.0}
\plotone{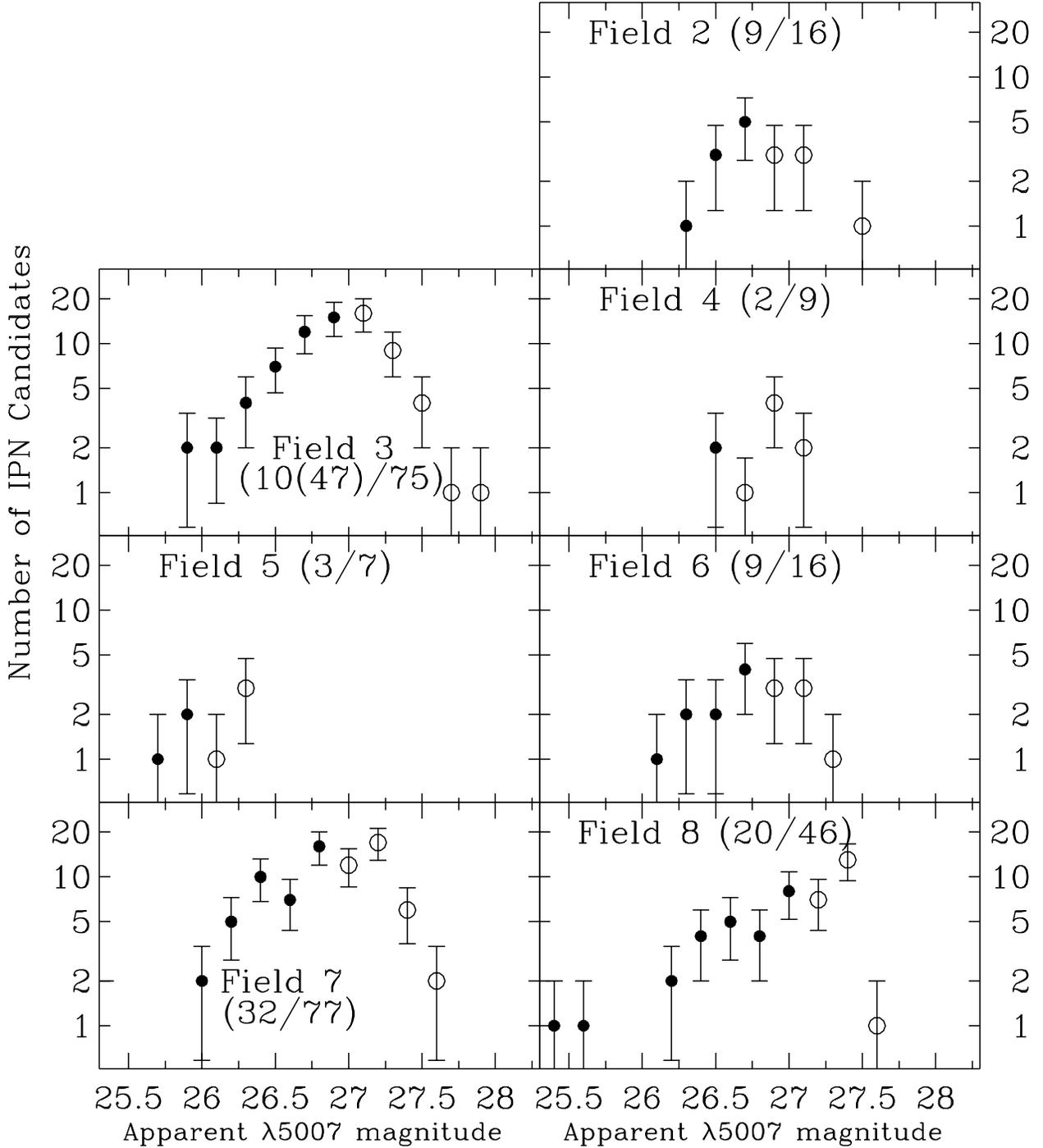}
\caption{The luminosity functions of our IPN candidates, binned into
0.2 magnitude intervals.  Filled points indicate objects brighter
than our derived completeness limits from artificial star experiments, 
and open circles indicate objects fainter than the completeness limits.
The numbers adjacent to each field show the number of IPN candidates above
the completeness limits, and the total number of candidates.
}
\label{fig:pnlf}
\end{figure}

After identifying the IPN candidates, we determined \oiii~magnitudes 
by comparing their fluxes to spectrophotometric
standards defined by Stone (1977) and Massey \etal (1988). We 
define our \oiii~magnitudes in the same way as Jacoby (1989): 
\begin{equation}
 m_{\rm 5007} = -2.5 \log F_{5007} - 13.74
\end{equation}
where $F_{5007}$ is in ergs~cm$^{-2}$~s$^{-1}$.
Equatorial coordinates were obtained for each intracluster planetary 
nebula candidate by comparing its position to those of USNO-A 2.0
astrometric stars (Monet 1998; Monet \etal 1996) on the same
frame.  The magnitude limit of each frame was determined by adding
artificial stars to the data and estimating the recovery fraction
as a function of magnitude (\eg~Stetson 1987; 
Bolte 1989; Harris 1990). To be conservative, the 90\% 
completeness level has been adopted as the limiting magnitude for each field.
The field locations are shown graphically in Paper~II, Figure 1.  
The resulting luminosity functions for each field, binned into 0.2 magnitude
intervals, are shown in Figure~\ref{fig:pnlf}.  The original data for
Field~1 has been superseded by the Field~7 results (Paper~II).    

\section{The Spatial Distribution of Intracluster Starlight}

The first step towards understanding the mechanisms which produce 
intracluster stars is to compare the stars' spatial distribution with that
of the better-studied components of galaxy clusters: 
the cluster galaxies, the hot intracluster gas, and 
the invisible dark matter, which 
dominates the total mass.  For example, if most intracluster stars are
removed early in a cluster's lifetime (Merritt 1984), the distribution
of this diffuse component will be smooth and follow the cluster potential.  
However, if a significant portion of the stars are removed at late epochs via 
galaxy encounters and tidal stripping (\eg~Richstone \& Malumuth 1983; 
Moore \etal 1996), then
the ICL should be clumpy and have a non-relaxed appearance.  In particular,
the ``harassment'' models of Moore \etal (1996) and Moore, Lake, 
\& Katz (1998) first predicted
that many intracluster stars will exist in long ($\sim 2$~Mpc) tidal
tails, which may maintain their structure for Gigayears.  Since Virgo is
thought to be dynamically young (Tully \& Shaya 1984; Binggelli, Popescu 
\& Tammann 1993 and references within), it is possible that
such structures are still present and observable.  

Since planetary nebulae closely follow the distribution of starlight in
galaxies (\eg~Ciardullo, Jacoby, \& Ford 1989; Ciardullo \etal 1989), 
they should be an ideal tracer of Virgo's intracluster population.
Here, we detail our efforts to determine the spatial distribution of 
the intracluster stars through the analysis of the IPN data.

\subsection{The Depth of Intracluster Starlight in the Virgo cluster}

From theory (\eg~Dopita \etal 1992; Stanghellini 1995; M\'endez \& 
Soffner 1997), and observations (\eg~Ciardullo \etal 2002b, and references 
therein), it has been shown that bright planetary nebulae follow a 
well-defined luminosity function in the light of \oiii.  In particular,
the [O~III] planetary nebula luminosity function (PNLF) at the brightest
magnitudes is well fit by an exponential function with a sharp 
bright-end cutoff $M^*$, \ie 
\begin{equation}
N(M) \propto e^{0.307 M} \{ 1 - e^{3 (M^* - M)} \}
\end{equation}
Since there is no reason to believe that IPNe would not 
also follow the PNLF, it is possible to use this cutoff to 
probe distances in the intracluster environment.  However, an
implicit assumption of the PNLF method is that all the PNe that
make up the observed luminosity function are effectively 
at the same distance (\eg~in a normal galaxy).
In a cluster such as Virgo, the finite depth of the cluster 
can distort the 
PNLF, as IPNe at many different distances can contribute to the observed 
luminosity function.  In this case, the PNLF of the IPNe 
will be the convolution of the intrinsic luminosity function 
and the density function of the IPNe along the line-of-sight.  
Thus, the shape of the observed luminosity function can be exploited to 
learn about the three-dimensional distribution of the ICL.

Unfortunately, deriving the complete line-of-sight distribution
is not trivial.  Mathematically, the function is:
\begin{equation}
O(m) = [(\rho(\mu) * N(M)) + C (m)] \ast E(m)
\end{equation}
where $O(m)$ is the final observed luminosity function, $N(M)$ is the
empirical PNLF, $\rho(\mu)$ is the density of
intracluster PNe as a function of distance modulus, $\mu$, $C(m)$ is the
apparent luminosity function of any contaminating objects, 
and $E(m)$ is the photometric error function of the 
observations.  Although the intrinsic luminosity 
function is known and the error function is, in principle, derivable from 
our photometry 
\citep[this function is not necessarily a Gaussian;][]{sh1988}, 
the luminosity
function of the contaminants is not currently not well-known.  This
places a fundamental limit on our ability to measure $\rho(\mu)$.  With an
improved estimate of $C(m)$, we will be able to do a much better job at
constraining this function.

For the present, we focus on a much simpler case: the magnitude
of the brightest IPN candidate in each field.  Given the empirical
PNLF, there is a maximum absolute magnitude $M^*$, that a PNe may
have in the light of \oiii.  If we assume that the brightest 
observed IPN candidate in each field has this absolute magnitude,
we can derive its upper limit distance, and therefore estimate
the distance to the front edge of the Virgo intracluster
population.  For this paper, we adopt a value of $M^*$ of
-4.51 (Ciardullo \etal 2002b). 

\begin{figure}
\epsscale{1.0}
\figurenum{2}
\plotone{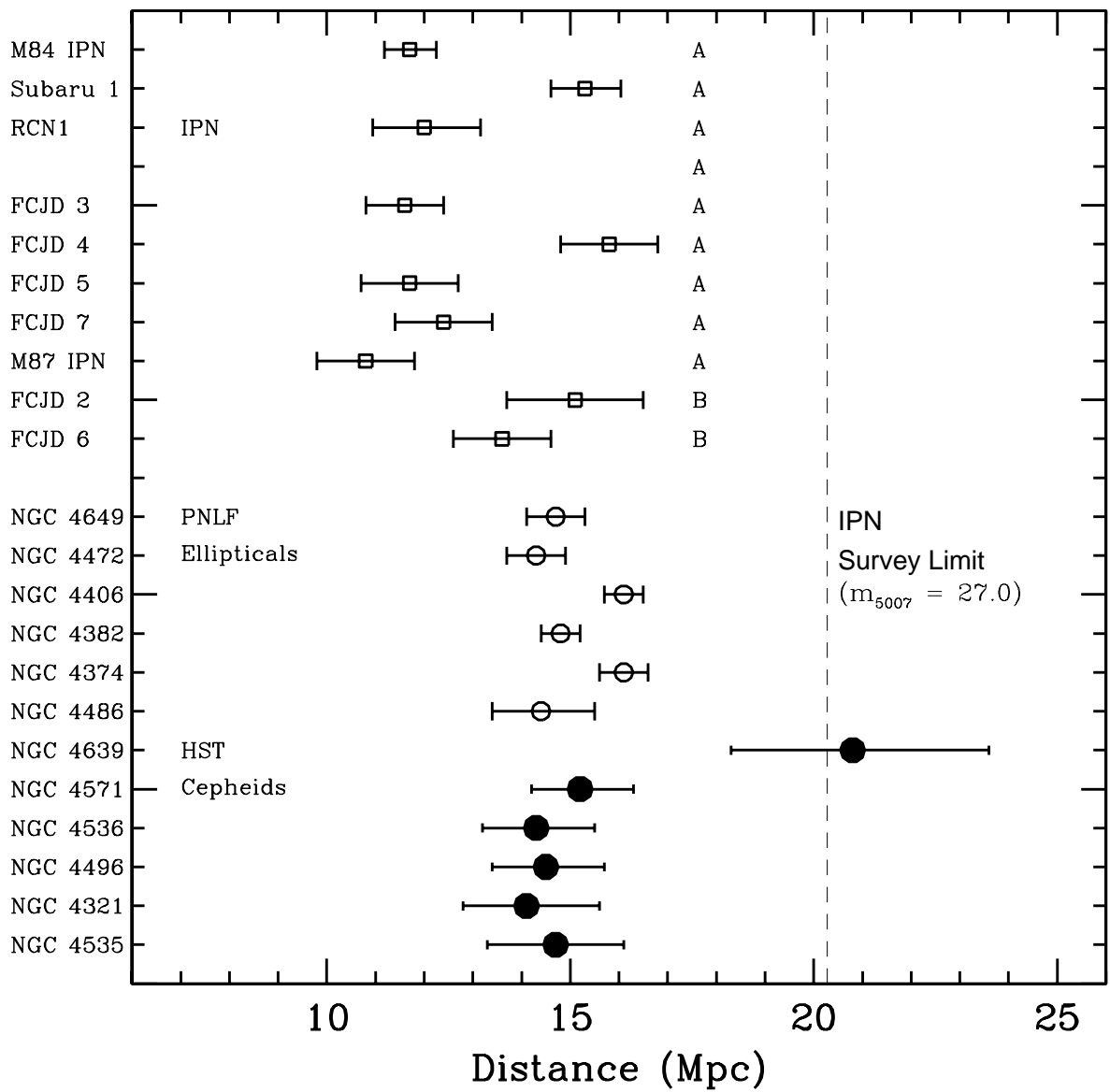}
\caption{A comparison of the upper limit distances obtained from the  
intracluster planetary nebulae to direct distances to Virgo
Cluster galaxies.  At the top are the upper limit distances (denoted 
by the open
squares) from IPN observations
by Okamura \etal (2002) and Arnaboldi \etal (2002).
Below that are the distances derived from the fields in Paper~II,
as well as the overluminous IPNe found in front of M87 (Ciardullo \etal 
1998).  These upper limit distances are compared to the PNLF distances of 
Virgo ellipticals (denoted by the open circles; 
Jacoby, Ciardullo, \& Ford 1990; Ciardullo \etal 1998), 
and Cepheid distances to spiral galaxies 
(denoted by the filled circles; Pierce \etal 1994; Saha \etal 1997; 
Freedman \etal 2001).  The subcluster of Virgo that each 
intracluster field resides in is noted.  See the text for 
further discussion.
}
\label{fig:depth}
\end{figure}

Figure~\ref{fig:depth} displays our upper limit distances, 
along with similar limits found by other authors 
using similar techniques \citep{m87ipn,rcn1,okamura2002}. 
Field~8 has been excluded from the 
analysis, since the IPN density of the field is consistent with zero 
(Paper~II).  The extraordinary IPN candidate I-33 of 
\citet{okamura2002}, which is over 0.8 magnitudes brighter than all
other objects in our survey, 
has also been omitted from the diagram.  For comparison,
Figure~\ref{fig:depth} 
also shows the Cepheid-based distances to six Virgo spirals 
\citep{pierce1994,saha1997,ceph2001}
and the PNLF distances of six Virgo early-type galaxies \citep{pnlf5,m87ipn}. 
Note that Virgo is known to have a complex structure with at least two
main subclusters: cluster~A, which is dominated by the giant elliptical
M87 (and contains the background M84/M86 subgroup;
B\"ohringer \etal 1994; Schindler, Binggeli, \& B{\" o}hringer 1999), and
cluster~B, which contains the giant elliptical M49 
\citep[][]{vc6}.  The respective subclusters for each field 
are labeled in the figure.

Several results are apparent from the data.  First, there is a 
general offset in the upper limit distances between IPN observations
in subclusters A and B.  According to our data, the front of 
subcluster~A is at a distance of 12.7 $\pm$ 0.4~Mpc, while
subcluster~B begins at 14.1 $\pm$ 0.8~Mpc.  These distance estimates use
all the datasets and include all possible systematic
errors common to the PNLF method (see Ciardullo \etal
2002b for a discussion of these systematic effects).  
If we take an average of the brightest  
$\rm m_{5007}$ magnitudes of our fields (2--7, and the M87 
field of Ciardullo et al. 1998) weighted solely by the photometric errors,
we find that the average magnitude of the brightest IPNe candidates
in subcluster A is $m_{5007} = 25.87 \pm 0.03$, while that of the 
brightest IPNe in subcluster B is $m_{5007} = 26.20 \pm 0.06$, corresponding 
to a difference at the 4.9$\sigma$ level.  If, instead of using 
just the brightest IPNe, we fit
the data to the empirical PNLF using the method of maximum likelihood
(Ciardullo \etal 1989), then the upper limit distance to each field
decreases by $\sim 0.7$~Mpc (Field~4 excepted, since that region does not
contain enough planetaries for a reliable PNLF analysis).  This systematic
shortening of the distances is expected, since it is more probable that
the brightest IPNe have magnitudes slightly fainter than $M^*$, rather
than precisely at $M^*$. Nevertheless, the analysis confirms the fact that
intracluster stars of subclump~B is substantially more distant than those
of subclump~A.

The systematic difference between the upper limit distances is most
likely due to the internal geometry of Virgo:  observations 
by Yasuda, Fukugita, \& Okamura (1997) and Federspiel, Tammann, 
\& Sandage (1998) demonstrate that the galaxies of subcluster~B have a 
mean distance modulus $\sim 0.4$~mag more distant than those of 
subcluster A.  Moreover, the two discrepant points in this relation, Field~4
and Subaru~1, can be easily explained through other means.  Field~4
only has two objects in its photometrically complete sample, so small
number statistics affect its distance determination.  (The probability
of one of these PNe being within $\sim 0.2$~mag of $M^*$ is only 
$\sim 30$\%.)  Subaru~1, on the other hand, is centered on the M84/M86 
subclump of cluster A, and this system is known to lie $\sim 0.2$~mag
behind the Virgo core (\eg~Binggeli, Popescu, \& Tammann 1993; 
B\"ohringer \etal 1994; Schindler, Binggeli, \& B{\" o}hringer 1999; 
Jerjen, Binggeli \& Barazzi 2003).  Therefore, the upper limit 
distance of Subaru~1 probably reflects the distance to this 
background subgroup, and not the Virgo cluster as a whole.  

A second property which is obvious from Figure~\ref{fig:depth} 
is that the
intracluster distance to subcluster~A is clearly much closer
than the 15~Mpc assumed for the cluster core.  
As has been pointed out previously (Ciardullo \etal 1998; 
Paper~I; Arnaboldi \etal 2002), this is due to a selection effect: 
it is much easier to detect
an IPNe located on the front edge of the cluster than one located on the
back edge.  Thus, our upper limit IPN distances represent measurements to the
front edge of Virgo's ICL, not the system's mean distance.  Nevertheless,
the depth implied from our measurements is remarkable.  
If we take the data at face value, then the IPN distribution 
has a line-of-sight radius of 4.2 + (15 * (D$_{15}$ - 1)) Mpc.  
If we compare this radius to the classical radius of Virgo on the sky 
\citep[$\sim 6\fdg 0$, or $\sim 1.6 D_{15}$~Mpc;][]{shap1926,gdv1973,vc6}
we find that the Virgo cluster is more than 2.6 times as deep as 
it is wide.  Virgo is nowhere near a spherical cluster: it contains 
considerable substructure, and is elongated significantly along 
our line of sight.

Our inferred depth for the Virgo Cluster is much greater than the
$\sim 2$~Mpc value associated with the early-type galaxies of the cluster's
core \citep[\ie][]{pnlf5, ton01, neilsen2000}.  On the other hand, this depth
agrees extremely well with that derived from Tully-Fisher observations of
the cluster's spirals.  According to \citet{y97} and \citet{sol2002},
Virgo is highly elongated along the line-of-sight, with a width-to-depth
ratio of $\sim 1$:4, and an overall depth of $\sim 10$~Mpc.  Much of this
depth lies background to the core ellipticals, but the line of spirals
does extend into the foreground.  This conclusion is strengthened by the
discovery of hydrogen deficient spirals on the outskirts ($R > 4$~Mpc) of
Virgo, which have likely moved through the cluster core \citep{sanchis2002}.
Moreover, it is also in agreement with the results of \citet{jerjen2003}, 
who find a Virgo Cluster depth of 6~Mpc from SBF measurements of its
dwarf ellipticals.  Since simulations 
\citep{mlk1998, dub2000} show that the majority of
intracluster stars are ejected into orbits similar to that of their parent
galaxies, our measurements imply that the bulk of Virgo's intracluster stars
originate in late-type galaxies and dwarfs.  If IPNe were derived 
from elliptical galaxies, the depth of the ICL would be much less.

A complication from this analysis is that we might be 
inadvertently combining two different components of ICL together 
in our analysis: a ``core'' 
component associated with the elliptical galaxies, and a 
component of stars removed from galaxy-galaxy interactions in the 
infalling regions of Virgo.  We believe this to be unlikely 
for a number of reasons.  First, modern models of intracluster star
production imply that the bulk of intracluster stars are
released when their parent galaxies and groups pass through 
pericenter (see Mihos 2004a; Willman \etal 2004 for some graphical 
examples).  The relative amount of intracluster star production is
much less on the outskirts of a galaxy cluster.  Second, if such
a component existed, the lower relative velocities of such encounters
would imply that there would still be luminous ``parent'' galaxies
nearby our IPN fields (100-200 kpc) in the foreground, assuming that the 
interactions happened 1-2 Gyr ago.  This does not appear to be the case 
(see \S 4.2).  

Finally, there is independent evidence that galaxies that have traveled
through the cluster core can return to radial distances of $\sim 4$~Mpc.
As mentioned previously, observations of H~I deficient spiral galaxies
show depths that are even larger that those derived from the IPNe
\citep{sanchis2002}.  Since the accepted explanation for H~I 
deficiency is ram pressure
stripping by the hot gas in the cluster core \citep{gunngott1972}, the
existence of these galaxies imply that at least some Virgo objects have
extended radial orbits.  Moreover, although the apocenter distances of
these galaxies are somewhat uncertain, due to errors in their Tully-Fisher
distances and other effects \citep{sanchis2002}, there is good reason to
believe that a depth of $\sim 4$~Mpc is not unreasonable.  Using standard
cosmological simulations, \citet{mamon2004} found that the maximum 
rebound radius reached by infalling galaxies is roughly
 between 1 and 2.5 times the cluster virial radii.  
When they scaled this result to the Virgo
Cluster, \citet{mamon2004} found a corresponding rebound 
depth of 4.1~Mpc, remarkably
similar to that found for the IPN distribution. Although the precise
agreement between the infall models and the IPN distances is probably
fortuitous, it does suggest that our measured depth for the intracluster
stars is plausible.  Models by \citet{moore2004}, which predict a maximum
orbital radius of $\sim 3$~virial radii, further support our result.

ICL comparisons with the hot intracluster medium are less straightforward.  
Multiple studies with {\sl ROSAT} and
{\sl ASCA} \citep[][and references therein]
{rosatvirgo1994,sch1999,shibata2001}, demonstrate that  
the X-ray 
emission from Virgo's intracluster medium extends $\sim 4$ 
to 5 degrees ($\sim 1.2 D_{15}$ Mpc) from M87, 
and generally follows the galaxy distribution.  However, the
line-of-sight depth of this component is unclear.  
Although intracluster gas is dissipative, and 
the majority of clusters are only mildly aspherical in their 
X-ray distributions (Mohr \etal 1995), there are cases where 
the X-ray contours are considerably elongated (up to a factor of 
three in depth; Mohr \etal 1995).  Therefore, although 
the data suggest that the IPNe are more widely distributed in 
depth than the gaseous intracluster medium, this cannot be proven.    

Our conclusions above have two caveats.  First, 
from the spectroscopic observations of Kudritzki \etal (2000) and
Freeman \etal (2000), it is clear that there are contaminating 
objects as bright as $\rm m_{5007} = 25.8$ present within our 
Virgo survey data.  Therefore, the upper limits displayed
in Figure~\ref{fig:depth} could be affected by these interlopers.
However, there are a number of reasons to believe that our 
estimates are accurate.  The most powerful argument in support 
of our distance estimates is the magnitude offset between the 
results from subclusters~A and B{}.  It is difficult to
see how this difference could be caused by background 
contaminating objects, since to first order, these objects are 
scattered uniformly across the sky.  Second, given a contamination
probability of $\sim 20\%$ (Freeman \etal 2000; 
Ciardullo \etal 2002a), it is unlikely that 
the majority of the bright IPN candidates are background galaxies.  
Finally, the spectroscopic observations of Ciardullo \etal (2002a) 
definitively show that the brightest IPN candidates in front of 
M87 have \oiii~emission, and are not background objects.   
From these data, the ICL extends at least $\sim 2$~Mpc in front
of Virgo's central galaxy.  Similar results have been found 
by narrow-band H$\alpha$ imaging and limited spectroscopic follow-up by 
\citet{okamura2002} and \citet{2003arna}.

A more troubling issue concerns the possibility of intracluster
H~II regions.  Gerhard \etal (2002) recently discovered an isolated
H~II region in the outer halo of NGC~4388 or possibly 
in intracluster space.  Such objects would emit \oiii, and, if
smaller than $\sim 50$~pc in size, would be indistinguishable from 
IPNe without deep continuum images or Balmer-line
spectroscopy.   Fortunately, such objects are rare.  Even if these objects
are distributed throughout Virgo's intracluster space, then the density 
reported by Gerhard \etal (2002) implies a contamination fraction 
of only $\sim 3\%$.  Additionally, \citet{vollmer2003} claim that
the isolated H~II region discovered by Gerhard \etal (2002)
was formed from gas that was ram-pressure
stripped from NGC~4388, and now is falling back into the galaxy.
If this is correct, it is less likely that large numbers of 
isolated H~II regions will be found far away from their parent 
galaxies (however, see Ryan-Weber \etal 2003 for some examples).  
Spectroscopic observations of the brightest IPNe would remove 
any uncertainty.

\subsection{The Angular Distribution of the Intracluster Stars}

\begin{figure}
\epsscale{0.8}
\figurenum{3}
\plotone{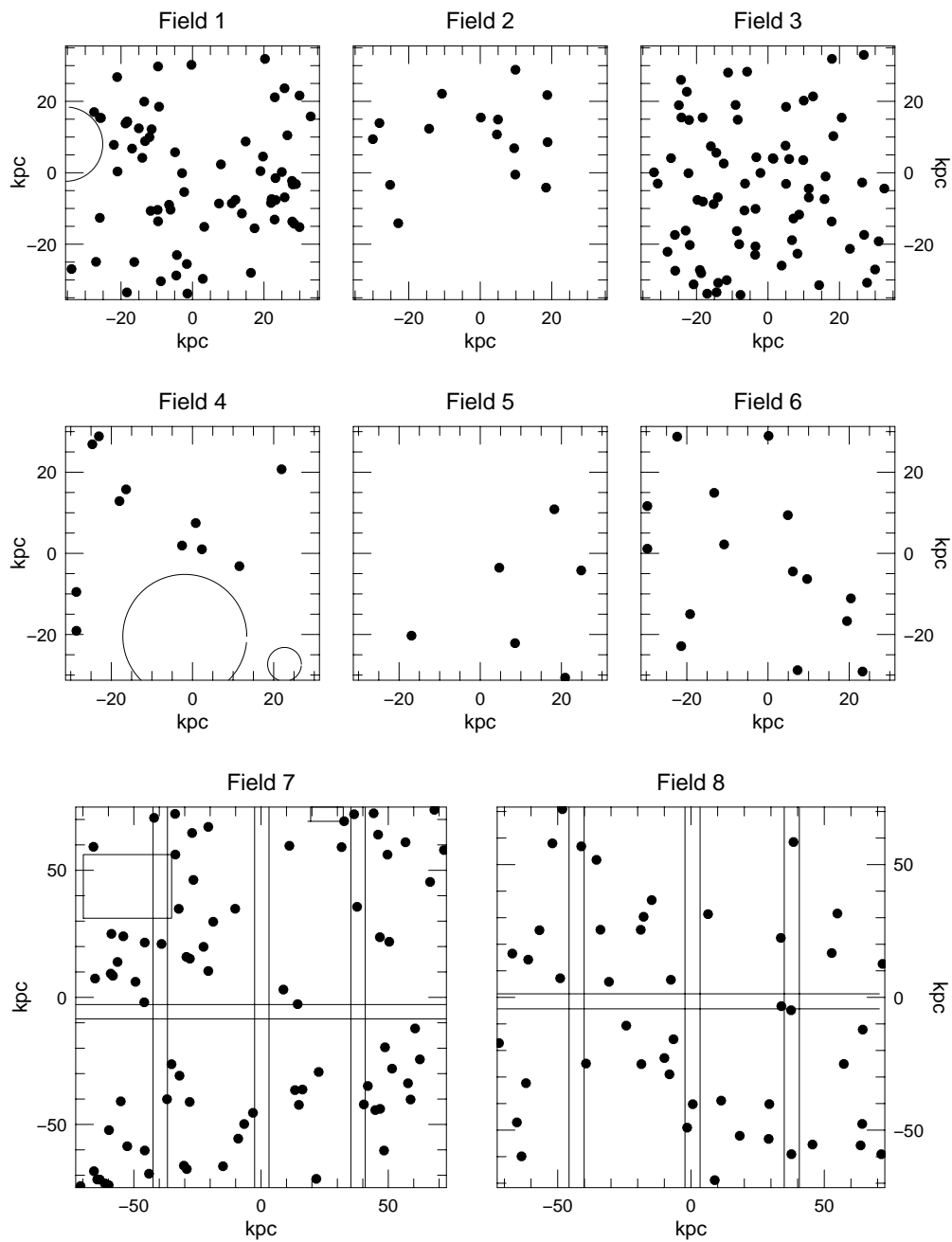}
\caption{The distribution of IPN candidates on the sky, denoted as the
filled circles.  Fields 1--6 are the PFCCD fields, and Fields 7--8 are the
MOSAIC fields.  Field 1 is included for completeness.  
Large regions that were excluded from each field are noted by circles
and rectangles.  These include the regions between the MOSAIC CCD chips that
are poorly surveyed and the areas around bright stars and galaxies.  
Note that there are no candidates in the bottom
third of Field 2, a feature not seen in other fields, and the general
clumpiness of the IPN distribution.
}
\label{fig:spatial}
\end{figure}

Another useful way to probe the distribution of Virgo's intracluster light
is through the locations of the IPNe on the sky.  These are shown in 
Figure~\ref{fig:spatial}.  
For reference, at the distance of Virgo, $1\arcmin$ corresponds to 
$\sim 4.36 D_{15}$~kpc.  Studying the IPNe in this way allows us to use
more of the IPN candidates in the analysis and investigate the 
intracluster light distribution on smaller spatial scales.  
It is obvious from the figure 
that at least some of our fields contain significant evidence 
of substructure.  For example, of the 16~IPNe of Field~2, 
none are present in the bottom third of the CCD frame.
This segregation is not instrumental -- none of the other fields exhibit 
that property.  From simple binomial statistics, the probability of the
bottom third of the field being empty is less than $\sim 0.15\%$.

Of course, the eye can be fooled, and {\it a posteriori} statistics are of
questionable validity.  What is needed is an unbiased statistical test of
the spatial distribution of IPN candidates.  The traditional approach
for such an analysis is the two-point angular correlation 
function (Peebles 1980), 
which is defined as the joint probability of finding two objects 
in two particular elements of solid angle on the sky.  Since 
the accuracy of this method depends on the number of object pairs 
in the sample, the limited of IPN candidates in Fields 2--6 
make this statistic too noisy for diagnostic purposes.

A simpler procedure that requires less data is quadrat 
analysis (see Diggle 1983 and references therein).  
This method is a two-dimensional analog to the more 
common three dimensional ``counts-in-cells'' 
test used in cosmological surveys (Peebles 1980), and 
was first used in studies of galaxy clusters by \citet{abell1961}.
Like the two point correlation function, quadrat analysis 
compares the observed IPN distribution against a random distribution.   
Unlike correlation functions, however, quadrat analysis has  
the ability, at least in theory, to test for the presence of 
sheets and filaments.

The principle of quadrat analysis is simple.  If one divides a field 
into $m$ regions, and the objects are distributed randomly, 
then the number of objects present within each region should 
be distributed as the $\chi^2$ statistic,

\begin{equation}
\chi^{2} = \sum^{m}_{i=1}  ( n_{i} - n_{\rm i, expected})^{2} / 
n_{\rm i, expected}
\end{equation}
where $n_{i}$ is the observed number of objects in region $i$ and
$n_{i,{\rm expected}}$ is the mean surface density of objects multiplied by the
region's area.  Note that there are no constraints on the size of each 
region, nor its shape.  However, given the limited number of IPN candidates
present in our data, we chose to simply divide our CCD fields into
rectangles and test for large-scale non-uniformity.

To implement our test, we began by dividing each CCD field into $2\times 2$,
$3 \times 3$, and $4 \times 4$ rectangular regions, and calculating the
effective area of each region.  Note that this second step was not
straightforward:  bright stars, galaxies, bad pixels, and gaps between the 
CCDs in our MOSAIC fields (Fields 7 and 8) all act to limit our IPN survey 
area.  To account for these effects, we very carefully masked out all those 
regions where the IPN detections would be abnormally difficult (or impossible),
and restricted our analyses to the areas outside these regions.

Our next step was to choose samples of IPNe for analysis.  Our IPN detections
on a frame are 90\% complete down to a limiting magnitude of $m_{\rm lim}$;
if we restrict our study to objects brighter than this, then we can be
assured that our quadrat analysis will not be biased.  However, it is possible
(and even likely) that the probability of detecting an IPN below the
completeness limit is independent of frame position, and only a function of
shot noise.  If this is the case, then we can use our entire sample of 
IPNe (which is typically a factor of $\sim 2$~larger than the statistically
complete sample), and still obtain a result that is bias-free.   To test
for this possibility, we performed our quadrat analysis on both the
statistically complete sample, and on our entire sample of IPN candidates.
For Fields 4 and 5 however, there are simply too few objects to measure
the spatial distribution for the statistically complete sample.

Tables~\ref{table:quad-all} and \ref{table:quad-comp}  
contain the results of our analysis.  From the table, it is
clear that the IPNe in several of our fields are non-randomly distributed.
The best evidence for this effect occurs in Field~7, where the statistically
complete sample of IPNe are non-randomly distributed at the 95\% confidence 
level, and the entire sample of IPNe are non-random with 99.9\% confidence.
Fields~2 and 4 also show some evidence for clustering, although the 
signal is weaker.  When their entire IPN samples in these fields are used,
random distributions are excluded with 95\% confidence.

An examination of Figure~\ref{fig:spatial} 
reveals that the features found by the quadrat 
analysis are easily seen by eye.  The signal for Field~2 comes from the
absence of objects in the lower portion of the frame, while Field~4's 
non-random result is generated by the cluster of objects in the upper
left-hand corner.  Finally, the large $\chi^2$ of Field~7 comes from
the east-west division of IPNe and the two large areas where IPN candidates
are seen.   These non-uniformities range in scale from $1\farcm 0$ to
$9\farcm 0$, and are in agreement with 
 the conclusions of Okamura \etal (2002; see
their Figure~1)  who find an inhomogeneous distribution of IPNe in 
their Subaru~1 field.  

How do these results compare with the ``Harassment'' scenarios of 
Moore \etal (1996)?  By both visual inspection of the Moore \etal
(1996) simulations, and order-of-magnitude arguments, the width
of tidal debris arcs should be $\sim$ 50~kpc, which corresponds 
to $\sim$ 12$\arcmin~\rm~D_{15}^{-1}$ in the Virgo cluster.  Observations 
of such arcs in the Coma and Centaurus clusters (Trentham \& Mobasher 1998; 
Gregg \& West 1998; Calca\'neo-Rold\'an, \etal 2000), confirm that
tidal tails are often $\sim 5$ to $\sim 40$~kpc in width
($1~-~10\parcmin0\rm~D_{15}^{-1}$).  This implies that a 
tidal debris arc would fill a significant 
portion of a PFCCD frame, and up to one one-third of a Mosaic field.  
Therefore, the non-uniformities detected in our analysis 
are more likely to indicate small scale inhomogeneities, either
within a longer tidal arc, or possibly 
unrelated to arcs.  More wide-field data will be needed to 
unambiguously search for long thin tidal debris arcs in the Virgo cluster. 

We note that since the density of contaminating sources is approximately 20\%, 
it is unlikely that they could have produced the signal found
by our quadrat analysis.  The evidence from both Lyman-Break galaxy surveys
\citep{adelberger1998,gia2001}, 
and higher redshift Lyman-$\alpha$ galaxy surveys (Ouchi \etal
2003), suggests that high redshift galaxies cluster on somewhat 
smaller angular scales ($\approx 2-3\arcmin$) than seen in our
data.  Still, the best way to confirm our clustering results 
is to obtain velocities for the IPN candidates.  
If these objects are part of a tidal debris arc, or
other related structure, they will be strongly correlated in velocity, 
as well as position.  Adding this third dimension is crucial to 
the interpretation of these data.

\section{The Amount of Intracluster Starlight}

In principle, determining the amount of intracluster luminosity from 
the observed numbers of IPNe is straightforward.  Theories of simple 
stellar populations (\eg~Renzini \& Buzzoni 1986) have shown that the 
bolometric luminosity-specific stellar evolutionary flux 
of non-star-forming stellar populations should 
be $\sim 2 \times 10^{-11}$~stars-yr$^{-1}$-$L_{\odot}^{-1}$, (nearly) 
independent of population age or initial mass function.  If the 
lifetime of the planetary nebula stage is $\sim 25,000$~yr (Kwok 2000),
and if the PNLF of equation (2) is valid to $\sim 8$~mag below
the PNLF cutoff,  
then every stellar system should have $\alpha \sim 50 \times 
10^{-8}$~PN-$L_{\odot}^{-1}$.  According to equation (2),
approximately one out of ten of these PNe will be within
2.5~mag of $M^*$.  Thus, under the above assumptions,
most stellar populations should have   
$\alpha_{2.5} \sim 50 \times 10^{-9}$~PN-$L_{\odot}^{-1}$.  
The observed number of IPNe, coupled with equation (2), 
can therefore be used to deduce the total luminosity
of the underlying stellar population.

Unfortunately, there are a number of systematic uncertainties 
associated with this conversion that must be addressed 
before we can use IPN surface densities to estimate 
the intracluster luminosity.  We discuss each of these
uncertainties below and determine a conservative 
correction for each effect.  

\subsection{Systematic Uncertainties}
%Alpha results
Although stellar evolution predicts a constant $\alpha_{2.5}$
value for all non star-forming populations, observations present a 
more complicated picture.  Ciardullo (1995) found that in a sample 
of 23 elliptical galaxies, lenticular galaxies, and spiral bulges, 
the observed value of $\alpha_{2.5}$ never exceeded 
$\alpha_{2.5} = 50 \times 10^{-9}$~PN-$L_{\odot}^{-1}$ but was 
often significantly less, with higher luminosity galaxies 
having systematically smaller values of $\alpha_{2.5}$, down to values of   
$\alpha_{2.5} = 8.3 \pm 1.5 \times 10^{-9}
$ for M87 and $\alpha_{2.5} = 6.5 \pm 1.4 \times 10^{-9}$
for M49.  
A lower value for $\alpha_{2.5}$ has also been inferred for the populations
of Galactic globular clusters \citep[$\alpha_{2.5} \approx 13$;][]
{jacoby1997}.  The reason for this
discrepancy is unclear.  It is possible that in old or metal-rich populations,
not all stars go through the planetary nebula phase \citep{rbciau, jacoby1997};
such an hypothesis is supported by the anti-correlation between $\alpha_{2.5}$
and the presence of extreme horizontal branch stars in elliptical galaxies
\citep{ferguson1999}.  Alternatively, the PNLF may not always follow the simple
exponential law of equation (2).  Ciardullo \etal (2004) have recently 
shown that in
star-forming populations, the PNLF has a distinctive `dip' $\sim 2$~mag
below the PNLF cutoff; if similar features occur in older populations, then
the extrapolation to the 
faint end of the luminosity function may be inaccurate.
In either case, since the total amount of intracluster luminosity depends
on $\alpha_{2.5}$, the variation of $\alpha_{2.5}$ with stellar population
can be a significant source of error, especially considering that our
constraints on the age and metallicity of the stellar population are
still weak (Paper~I; Durrell \etal 2002).

To correct for this effect, we adopted the results of Durrell \etal (2002),
who used the {\sl Hubble Space Telescope} to measure the luminosity
function of red giants in Virgo's intracluster space.  Models 
demonstrate that these data can be easily scaled to give the total
luminosity of the intracluster population in this field
\citep[\eg][]{girardi2000}.  By comparing the numbers of IPNe in 
Field~7, which surrounds the {\sl HST}~field, with RGB and
AGB star counts, Durrell \etal (2002)
found a value of $\alpha_{2.5} = 23^{+10}_{-12} 
\times 10^{-9}$~PN-$L_{\odot}^{-1}$ 
for Virgo's intracluster population.  This value is similar to 
that derived for the bulges of spiral galaxies, and we adopt it
here for our luminosity calculations.  However, even this 
determination has a substantial systematic uncertainty.  As we
have noted in \S 3.2, the intracluster stars are not uniformly 
distributed on the sky, and the {\sl HST} survey field is only
$\sim 1/20$ the size of Field~7.  Thus there is the possibility 
of a density mismatch between the two measurements due 
to the spatial structure of the 
intracluster light.  At this time, we cannot determine the amount 
of this effect: additional intracluster red giant fields will be 
needed for a more accurate comparison.

A second systematic error associated with our measurements 
comes from the fact that samples of IPN candidates are not pristine.
Approximately 20\% of our IPN candidates are 
background objects of various kinds, of which the most common
are Lyman-$\alpha$ galaxies at z $\approx$ 3.12 
(\eg~Kudritzki \etal 2000; Rhoads \etal 2000; Stern \etal 2000).  
The only absolutely secure way to eliminate these contaminants is through
spectroscopy:  true IPNe have narrow ($\sim 30$~\kms) emission lines,
show additional flux at [O~III] $\lambda 4959$, and have an [O~III] $\lambda
5007$ to H$\alpha$ emission line ratio of $\gtrsim 2$ \citep{pnlf12}.  This
distinguishes IPNe from background galaxies (which have broader lines and
no companion emission) or intergalactic H~II regions (which usually have
an H$\alpha$ line that is brighter than [O~III] $\lambda 5007$).
Unfortunately, since we do not yet have follow-up spectroscopy for our
IPN candidates, we must statistically remove the background contaminants
using the methods detailed in Paper~II and in Ciardullo \etal 
(2002a).  Using a set of control fields, and identical 
search procedures as the Virgo IPN survey, we determined the 
surface density of 
IPN-like objects, and adopted that density as our mean background 
(69$\pm$ 23 per square degree brighter than $m_{5007} = 27.0$).
This density is corrected in each field for any known background
objects (Paper~II).

There are a number of complications to this background subtraction.
First, as noted in Paper~II, the Ciardullo \etal (2002a) survey
only covered 0.13~deg$^{-2}$ of sky, and thus detected only nine
unresolved IPNe-like objects.  Consequently, the estimate of 
background density has a large Poissonian error.  Of 
greater concern is the effect that cosmic large-scale structure
has on the background source counts.    
Recent observations by Ouchi \etal (2003) find that Lyman-$\alpha$ 
galaxies are already clustered at z=4.86, which implies even 
greater clustering at z $\approx$ 3.  Combined with the z $\approx 3$ 
Lyman-break galaxy results of \citet{adelberger1998} and \citet{gia2001}, 
it is almost certain that our background estimation is affected by
the large-scale structure exhibited by emission-line galaxies.
We do note that our contamination fraction
is similar to the $\sim 26\%$ value found by Freeman \etal (2000) 
from limited follow-up spectroscopy and the $\sim 15\%$ estimate
of \citet{castro2003} based on a narrow-band survey of a field
in the Leo~I galaxy group.  Ultimately additional blank field 
measurements and spectroscopic follow-ups will be needed to 
further reduce the uncertainty.

Finally, as noted in \S 3.2, the IPN candidates of Virgo have a 
significant line-of-sight depth.  Because our observations do 
not reach all the way through the cluster, our density 
determination is biased: 
we are more likely to detect IPNe on the near side of the 
cluster than the far side.  In Paper~I, we created two 
extreme models for the IPN distribution: a minimal ICL model, 
wherein all the IPNe are at a single distance, and a maximum
ICL model, in which the intracluster stars are distributed 
isotropically throughout a sphere of radius 3~Mpc. 
The difference in densities between these two models 
is a factor of $\sim 3$.  Therefore, we are faced with a 
quandary: do we assume a spherically 
symmetric model, and multiply the measured densities by a factor of 
three, or do we adopt the single distance model that we know to be incorrect?
In order to be conservative, we have chosen the latter option, and
adopt a single distance model for all of our intracluster fields.  
In the future, when our studies of the line-of-sight densities 
are completed, we will be able to obtain a more accurate answer.  

In conclusion, although there are significant systematic uncertainties
relating the number of IPNe to the total amount of starlight, we believe
we have reasonable corrections to these effects which are conservative,
and supported by the data.

\subsection{Fitting the PNLFs, and Determining the Amount of Intracluster 
Light}

To determine the amount of luminosity in each of our survey fields, we
began by adopting the PNLF of equation (2), and convolving it with the
photometric error versus magnitude relation derived from our photometry.
We then computed the likelihood of this model luminosity function fitting
the observed statistical sample of PNe as a function of two variables,
system distance and total PN population \citep{pnlf2}.  We then integrated over
the former variable, reduced the latter variable by the fraction of
contamination expected for each field, and then translated the PN populations
into intracluster luminosity via our best-fit value for $\alpha_{2.5}$.
These results appears in Table~\ref{table:luminosity}, column 2.  
The quoted errors are the formal uncertainties associated with the
maximum likelihood solutions; these are usually asymmetric 
due to the shape of the 
\oiii~planetary nebulae luminosity function.  
In the case of Field~3, we have shown previously that the PNe in 
this field are a combination of IPNe, and PNe in the
halo of M~87 proper, and that at least 10 of these objects 
can legitimately called IPNe (Paper~I).  
Therefore, to place limits on this field's
intracluster light, we assumed that 1) all
of the PNe candidates in this field are intracluster, or 2) that only
the brightest 10 candidates are intracluster.  These two models are
referred to as ``3h'' and ``3l'' respectively.  For Fields~4 and 5,
due to the very small number of objects above the completeness limit, 
the uncertainties are poorly constrained.  In these
cases, we replaced the maximum-likelihood error bars with the errors expected
from Poissonian statistics.  For Field~8, the IPN density is consistent with
zero, so no analysis has been performed in this region.

As expected, the amount of intracluster light is significant: the
intracluster luminosity contained within our average PFCCD survey
field is comparable to that of the Large Magellanic Cloud. 
The implied bolometric luminosity surface densities, denoted
as $\Sigma_{\mbox{IPN}}$, 
are given in Table~\ref{table:luminosity}, column~3.  For reference,
these densities are roughly $\sim 40$~times 
less than that of the Milky Way's disk at the solar neighborhood 
\citep{gdv1978}.  If we adopt a 
bolometric correction consistent with a late-type stellar population, 
($-0.80$), these bolometric luminosity surface 
densities can be translated into $V$-band surface
brightnesses.  These are given in 
Table~\ref{table:luminosity}, column~4.  Although the surface brightnesses
are quite faint (5--6 magnitudes below that of the night sky),
they are reachable with the newer generation of CCD ICL surveys now
being performed, provided that precise and 
accurate sky subtraction can be achieved (Gonzalez \etal 2000; 
Feldmeier \etal 2002; Krick \& Bernstein 2003; Mihos \etal 2004, 
in prep.).  

With the luminosity of Virgo's intracluster stars now determined, we can next 
compare this number to the amount of light contained within the cluster's 
galaxies.  This is non-trivial, because as we mentioned previously,
the structure of Virgo is unrelaxed, and difficult to model accurately.

To estimate this quantity, we used the galaxy catalog of \citet{vcc} to find 
the total $B$-band magnitudes of all member galaxies brighter than 
the catalog's limit of $B_T \sim 20$.  (This limit is adequate for 
our purpose, since the contribution of fainter
objects to Virgo's total luminosity is negligible.)  We then converted these
$B$-magnitudes to $I$, using the observed $B-I$ colors of early-type
galaxies \citep{goud94} and a mean $(B-I)$ color for late-type galaxies
\citep{dejong94}.  Finally, with the $I$ magnitudes of all the galaxies 
in hand, we computed the radial profile of Virgo in bolometric light,
using an $I$-band bolometric correction of $+0.4$.  Note that 
to do this,
we must determine the center of the cluster.  \citet{vc6} have 
shown that the luminosity density of Virgo peaks $\sim
58^\prime$ NW of the central elliptical, M87.  However, based on the cluster's
kinematics \citep{bin1993}, distances \citep{pnlf5, ton01, wb2000, fer00}, 
and X-ray
properties \citep{rosatvirgo1994,sch1999}, 
it is likely that this offset is due to the 
contribution of the M84/M86 Group, which is falling into Virgo from behind.
We therefore used M87 as the central point, and computed the 
$I$-band luminosity density of Virgo's galaxies in a series of 
circular apertures centered on the galaxy.  

The luminosity surface 
densities we derive are given in Table~\ref{table:luminosity},
column 5, which we denote as $\Sigma_{\mbox{galaxies}}$.  
For Fields 2 and 6, which are in subcluster B of Virgo, the circular
aperture model derived above is not applicable, since that 
subcluster has an irregular galaxy distribution.  
Therefore, to set a robust lower limit on the fraction derived
for Fields~2 and 6, we assumed a value for $\Sigma_{\mbox{galaxies}}$ 
that is representative of the center of subcluster A.

Finally, we computed the fraction of intracluster light in each field
by dividing $\Sigma_{\mbox{IPN}}$ by the sum of $\Sigma_{\mbox{galaxies}}$
and $\Sigma_{\mbox{IPN}}$.  At this point, to make error analysis 
more tractable, we also adopted the average uncertainty of 
our asymmetric density error bars.  Based on all of these assumptions, 
we derive intracluster star fractions between 4\% and 41\%, 
depending on the field.  When averaged over all our fields, the
fraction of intracluster light in Virgo is 18.5\% $\pm$ 3.0\% 
(excluding Field~8) and 15.8\% $\pm$ 3.0\% (with Field~8 included).  
The quoted uncertainties include only the random component; the
systematic uncertainties are dominated by the uncertain 
line-of-sight depth of the intracluster stars, and the error
associated with our estimate of $\alpha_{2.5}$.  We estimate
that these systematic errors could be as large as 5\%.
However, by assuming a single-limit distance for the intracluster
starlight, we have underestimated the total amount of intracluster
light, and therefore the fractions quoted above are conservative.

To ensure that our values for $\Sigma_{\mbox{galaxies}}$ are
reasonable, we have also calculated local luminosity surface densities for
each field to compare against our IPN luminosity densities.  
To do this, we defined two densities, $\Sigma_{25}$ and $\Sigma_{50}$,
to be the amount of bolometric luminosity density associated with the 25
and 50 \citet{vcc} galaxies closest to the center of each of our
survey fields, omitting clear background galaxies.  
We then used these densities to normalize our IPN
measurements in exactly the same way as described above.  The results are
given in Table~\ref{table:localluminosity}.  
For any individual field, there can be a large difference
between the locally and globally derived luminosity fractions.  This is
expected, given the clumpy, unrelaxed state of the cluster, and the
limited number of luminous galaxies near each field.  However, the
weighted mean of these two local determinations 
for the entire cluster is 15.7\% $\pm$ 1.4\% (including Field~8),
essentially identical to our previous result.

These fractions agree with single field studies of Virgo's
intracluster population.  Arnaboldi \etal (2002)
found a fraction of at least 17\% from the observations of field
RCN1, and Okamura \etal (2002), and Arnaboldi \etal (2003) 
estimated a fraction of 10\% from IPN data of the 
Subaru~1 field.  Similarly, Ferguson, Tanvir \& 
von Hippel (1998) found fractions of between 4\% and 12\% 
using {\sl HST\/} observations of IRG's, and 
Durrell \etal (2002) obtained a fraction of 15$^{+7}_{-5}$\% 
using the red giant measurements of two {\sl HST \/} fields.
However, it is important to note that
each of these studies makes different assumptions concerning
the calculation of intracluster luminosity, the amount of
light bound to galaxies, and the contamination of background sources.
So none of these numbers is unassailable.

How much stellar mass does this amount of intracluster light imply?
If we take our value for the median luminosity surface density of
intracluster light ($6.4 \times 10^5 L_{\odot}$~kpc$^{-2}$), and assume a
bolometric correction of $-0.80$ (a value typical of older stellar
populations) and a mass-to-light ratio of $M/L_V \sim 7.7$
\citep{buzzoni1989}, then the derived surface mass density of Virgo's
intracluster stars is $7.3 \times 10^{11} M_{\odot}$~Mpc$^{-2}$~D$_{15}^{-2}$.
Although this mass density seems large, when compared to a detailed
mass model of the three Virgo subclusters \citep{sch1999,sch2002},
and after correcting for the different assumed distances to the Virgo
core in the two calculations (15 Mpc versus 20 Mpc), we find that 
the intracluster stars add on average of $\approx$ 17\% to the stellar mass 
of the cluster.  In any case, as in most galaxy clusters, the majority of the
baryonic mass of Virgo is in the intracluster gas 
\citep{sch1999,sch2002}, and the baryonic mass itself is only a fraction
of the total mass.  Intracluster stars do not solve the missing-mass
problem in galaxy clusters, as was originally determined by \citet{gdv1960}.

\begin{figure}
\epsscale{1.0}
\figurenum{4}
\plotone{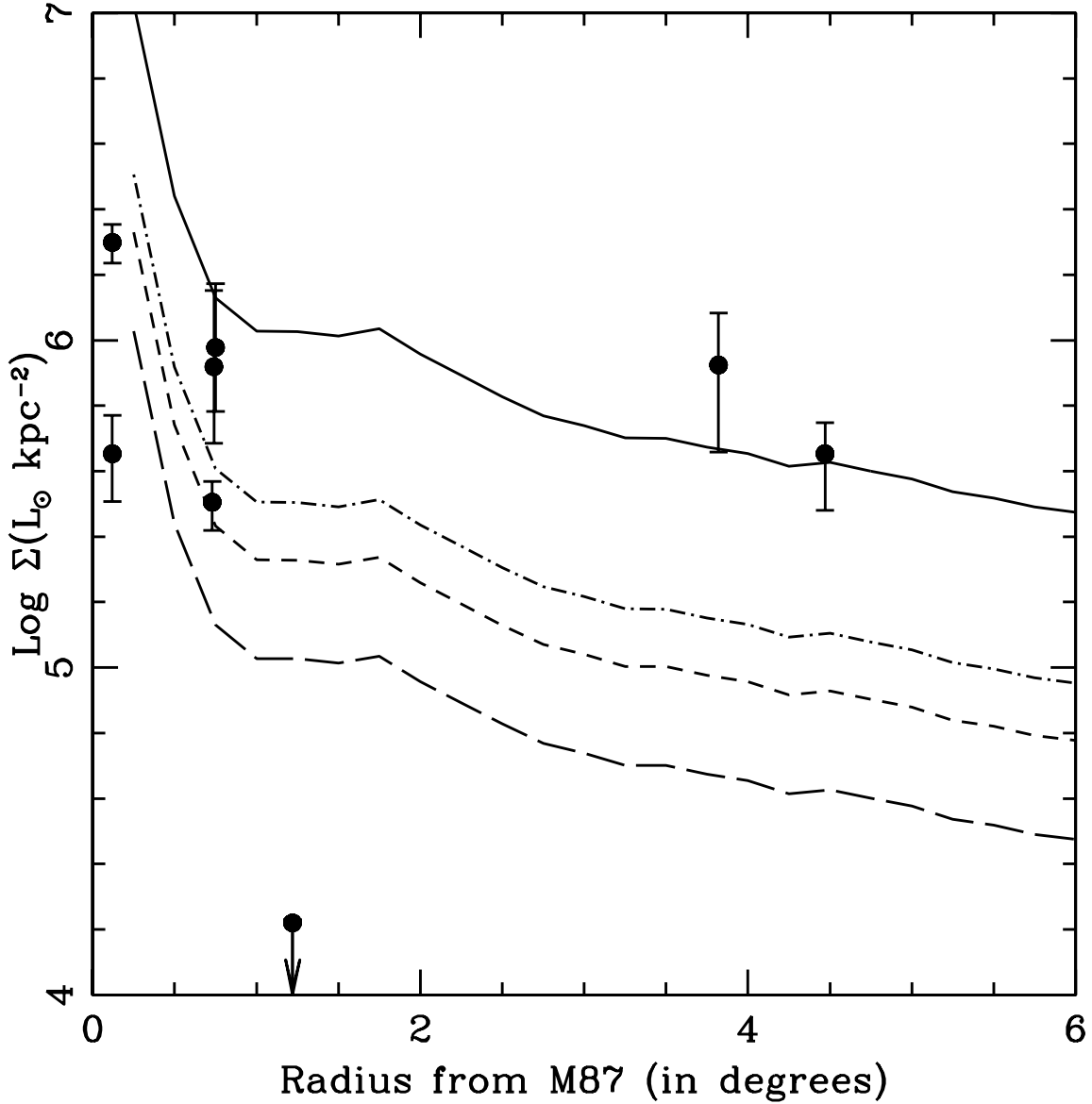}
\caption{The radial distribution of Virgo's intracluster light,
derived from the IPN observations, compared to galaxy luminosity 
density.  The solid line denotes the cluster's cumulative luminosity 
density (see the text for how this quantity was calculated).  The three
dashed lines from bottom to top show the 
10\%, 20\% and 30\% fractions, respectively.  Note that the decline of 
the IPN density is quite small compared to that
of the galaxy light.  Field~8, which has a density consistent with that
of the background is denoted by the upper limit at R $\approx1\degr$}
\label{fig:radial}
\end{figure}

The radial distribution of the IPNe compared to that of the cluster 
light is shown in Figure~\ref{fig:radial}.  As can be clearly seen,
the IPN density falls slowly, if it all, and certainly more slowly than
that of the galaxies.  If we fit the IPN densities with a 
simple power law in radius we find that the least-squares 
slope is $-0.1 \pm 0.5$.  Given the uncertainties associated
with the measurements, this result is marginally 
consistent with the simulations of Dubinski (1998), and
Napolitano (2003), which predict \rquart~like profiles for 
the intracluster light.

\begin{figure}
\epsscale{1.0}
\figurenum{5}
\plotone{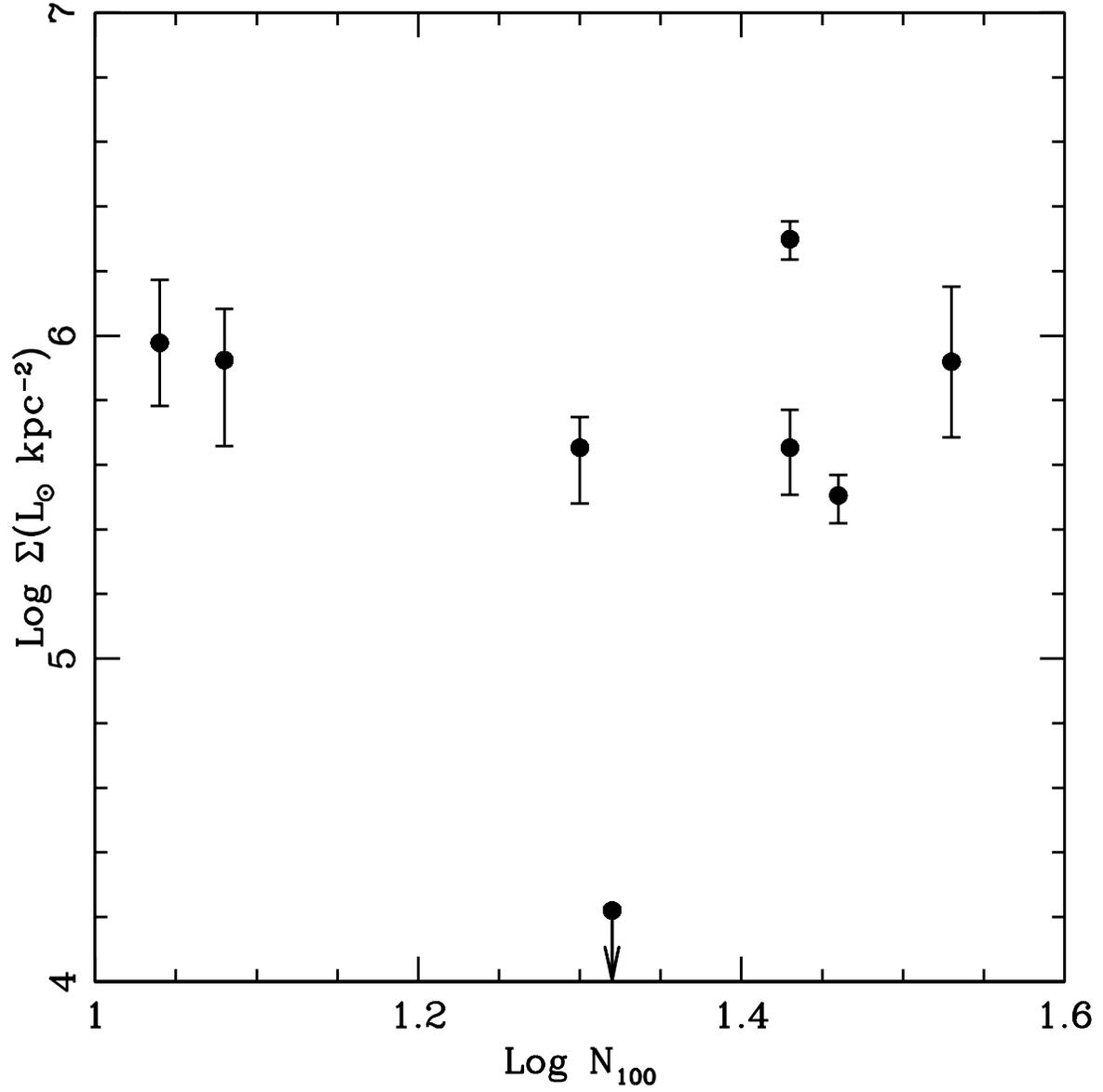}
\caption{The density of intracluster light in Virgo, compared to local
projected galaxy density within 100$D_{15}$~kpc of each field.  
There appears to be little or no correlation with galaxy projected 
density.  Field~8 has been given an upper limit, similar to Figure~4.}
\label{fig:dens}
\end{figure}

Since galaxy morphology correlates strongly with galactic
density \citep{dress1980}, it is conceivable that IPNe density
also correlates with this quantity.  To test this hypothesis, we 
obtained the local projected galaxy density for each field by taking the 
catalog of \citet{vcc}, and finding all Virgo galaxies within       
100$D_{15}$~kpc in radius ($23\farcm 3$) of each field center. 
By using this catalog, we are sensitive to 
objects at least as faint as $M_B \sim -13$, and in some cases,
two magnitudes fainter, which puts our density measurement into the 
domain of intermediate dwarf galaxies.  This number density, 
denoted as N$_{100}$, is given in Table~3, column 7, and plotted 
against the IPN density in Figure~\ref{fig:dens}.  

As can be clearly seen, there appears to be 
little or no correlation between intracluster light density 
and projected galaxy density.  After fitting the
data to a simple power-law, we find a slope of -0.1 $\pm 0.4$ in units
of log $\Sigma / \mbox{log} N_{100}$.  This null result is somewhat 
surprising, since if intracluster light is produced from galaxy
interactions, a correlation with density might be expected.  
We repeated this test over radii of 200, 500, and 
1000$D_{15}$~kpc respectively, and found similar null results 
(no positive slope at greater than the 2 $\sigma$ level).

These null results may be due to the significant scatter in the projected 
density, as well as the known uncertainties in the intracluster 
stellar densities.  Alternatively, the dynamic range of density surveyed 
in this analysis may be too small (a factor of $\sim 2$)
compared to that needed to see the morphology-density relation.
Finally, our measurements may suffer from an unavoidable selection 
effect.  
In order to avoid confusion with PNe which are bound in the halos of
galaxies, we must conduct our intracluster light survey in regions where
there are no bright objects.  In the past, these locations may have been
occupied by bright galaxies, and the intracluster stars may have been
released from these objects.  However, given the large amount of time that
could have elapsed between the release of these stars and the present
epoch, it is possible that the IPNe have drifted more than 
the adopted 100--1000$D_{15}$~kpc radii.  More intracluster 
observations over a larger range of cluster densities will be 
needed to address this issue further. 

\section{Discussion}

When we combine our own IPN data, the results from intracluster 
red giant observations \citep{ftv1998,durr2002}, 
and other IPN surveys \citep{1996arna,rcn1,okamura2002,2003arna}, 
we come to a reasonably consistent picture of Virgo's stellar 
intracluster population.  

There is a moderate (10--20\%) amount of intracluster light in the 
Virgo cluster, significantly more than found in the galaxy groups 
of M~81 ($<$ 3\%; Feldmeier \etal 2003b) and Leo~I ($1.6^{+3.4}_{-1.0}\%$;
Castro-Rodr{\'{\i}}guez \etal 2003), but much less than is present
in the rich clusters of Coma ($\sim$ 50\%; Bernstein \etal 1995), Abell~1689 ($\approx$ 30\%; Tyson \& Fischer 1995), and the
compact group HCG~90 \citep[38\%--48\%;][]{white2003}.
This is consistent with the view that Virgo is a dynamically young
cluster still in the process of formation \citep{tully1984,vc6,bin1993}.  
As Virgo dynamically evolves, the amount of intracluster light 
should increase.      

From the large depth derived from the IPNe, it seems clear that
a significant portion of Virgo's intracluster stars originate
from late-type galaxies 
whose highly radial orbits take them in and out of the cluster core.
The intermediate value of $\alpha_{2.5}$ also strongly implies that 
the bulk of the IPNe come from lower-luminosity galaxies, and 
not giant ellipticals (M$_{B} \geq -20$; Ciardullo 1995).
This conclusion is supported by the intracluster 
red giant star observations of
\citet{durr2002}.  Their analysis of the intracluster red giant 
luminosity function shows that
Virgo's intracluster stars have a mean metallicity in the range
$-0.8 < {\rm [Fe/H]} < -0.2$.  Such a value is more metal-rich than a 
typical dwarf galaxy \citep{shetrone2001}, yet more metal-poor than the
giant ellipticals that inhabit the Virgo cluster core 
\citep[i.e.,][]{terl1981,davies1987}.  Moreover, the value 
of $\alpha_{2.5}$ derived for the
IPN population, and the IPN properties described in \citet{m87ipn} and
Paper~I suggest that the intracluster stars of Virgo are not extremely
old or young.  If the IPN were extremely old, then the depth of the
Virgo cluster would be even higher than its large derived value (Paper~I),
and measurements of $\alpha_{2.5}$ in young stellar populations such
as the Magellanic Clouds give values near the theoretical maximum 
(Ciardullo \etal 1995).  These facts, along with the 
lack of correlation between IPN density and galactic 
environment, imply that the production of intracluster stars is 
an ongoing process.   

The above properties appear consistent with tidal-stripping from
galaxy interaction (Richstone \& Malumuth 1983) 
scenarios such as ``galaxy harassment''
(Moore \etal 1996; Moore, Lake, \& Katz 1998) ``tidal stirring''
(Mayer \etal 2001), and ``galaxy threshing'' (Bekki \etal 2003)
and are inconsistent with tidal stripping by the mean cluster field 
very early in the cluster's lifetime (Merritt 1984).  
However, a few caveats are in order.  First,
since we derive properties through the IPNe, any stellar
population that is extremely old and metal-poor (Jacoby \etal 1997), or
very metal-rich (Ciardullo 1995), will not produce large numbers of
IPNe, and so will be missed in our survey.  If there was a 
large amount of intracluster star production at early
times, it would be extremely old, and we would be insensitive
to this population.  The close agreement between the IRG and the 
IPN observations makes the existence of a large Pop~II/III 
component unlikely, but some extremely old intracluster stars 
may still exist.  Also, since IPNe are weighted by stellar 
luminosity, the properties of
rarer populations, such as those produced by tidally-stripped
dwarf galaxies, will be poorly represented in our survey.  
It is almost certain that 
some of Virgo's intracluster light originated from fainter 
dwarfs that have undergone severe tidal disruption and
now exist as a population of Ultra-Compact Dwarf galaxies 
(Drinkwater \etal 2003; Karick \etal 2003). 

There are a number of further studies that can improve the results
given in this paper.  Additional IPN fields, especially ones
at larger cluster radii and at lower galaxy densities, would be invaluable
in confirming our null results with respect to galaxy density and radial
distribution.  Deep imaging of a single field would
also assist in better defining the line-of-sight distribution of IPNe.
But the most important observations needed are follow-up spectroscopy 
for a large number of candidate IPNe.  Such observations would
remove the uncertainty in our results due to background sources, produce
better IPN magnitudes, and most critically, allow comparison of
Virgo's intracluster population with dynamical models.  

\acknowledgements

We thank M. Arnaboldi, K. Freeman, R. Kudritzki, R. M\'endez, 
and Chris Mihos for useful discussions, and their infinite patience.    
We also thank an anonymous referee for their comments which 
improved this paper.  J. Feldmeier acknowledges salary 
support from J.C. Mihos under grant AST-9876143.  This work 
was also partially supported by NSF grant AST 0071238, 
AST 0302030, and NASA grant NAG 5-9377.

\pagebreak

\begin{deluxetable}{ccccc}
\tablewidth{0pt}
\tablecaption{Quadrat Analysis of Total Sample
\label{table:quad-all}}
\tablehead{
\colhead{} & 
\colhead{} &
\colhead{} &
\colhead{} &
\colhead{Null Hypothesis} 
\\ \colhead{Field} & \colhead{Binning} 
& \colhead{$\chi^{2}$}
& \colhead{dof} &
\colhead{Probability}}
\startdata
2 & 2 & 7.48   & 3  & 0.06\\
2 & 3 & 10.9   & 8  & 0.21\\
2 & 4 & 26.3   & 15 & 0.04\\
\\
3 & 2 & 5.03   & 3  & 0.17\\
3 & 3 & 10.4   & 8  & 0.24\\
3 & 4 & 11.2   & 15 & 0.74\\
\\
4 & 2 & 2.73   & 3  & 0.43\\
4 & 3 & 14.2   & 7  & 0.05\\
4 & 4 & 20.2   & 15 & 0.17\\
\\
5 & 2 & 3.91   & 3  & 0.27\\
5 & 3 & 1.98   & 8  & 0.98\\
5 & 4 & 9.19   & 15 & 0.87\\
\\
6 & 2 & 4.08   & 3  & 0.25\\
6 & 3 & 6.51   & 8  & 0.59\\
6 & 4 & 12.1  & 15 & 0.67\\
\\
7 & 2 & 2.91   & 3  & 0.41\\
7 & 3 & 14.9  &  8  & 0.06\\
7 & 4 & 44.3  & 15 & $< 0.001$\\
\\
8 & 2 & 4.07   & 3  & 0.25\\
8 & 3 & 5.69   & 8  & 0.68\\
8 & 4 & 15.8   & 15 & 0.39\\
\enddata
\end{deluxetable}

\pagebreak
\begin{deluxetable}{ccccc}
\tablecaption{Quadrat Analysis of Photometrically Complete Sample
\label{table:quad-comp}}
\tablewidth{0pt}
\tablehead{
\colhead{} & 
\colhead{} &
\colhead{} &
\colhead{} &
\colhead{Null Hypothesis} 
\\ \colhead{Field} & \colhead{Binning} 
& \colhead{$\chi^{2}$}
& \colhead{dof} &
\colhead{Probability} \\}
\startdata
2 & 2 & 3.88 & 3  & 0.28\\
2 & 3 & 5.96 & 8  & 0.65\\
2 & 4 & 21.3 & 15 & 0.13\\
\\
3 & 2 & 6.34 & 3  & 0.10\\
3 & 3 & 12.3 & 8  & 0.14\\
3 & 4 & 14.2 & 15 & 0.51\\
\\
6 & 2 & 5.71 & 3  & 0.13\\
6 & 3 & 6.02 & 8  & 0.65\\
6 & 4 & 14.1 & 15 & 0.52\\
\\
7 & 2 & 1.29 & 3  & 0.73\\
7 & 3 & 7.24 & 8  & 0.51\\
7 & 4 & 25.2 & 15 & 0.05\\
\\
8 & 2 & 6.16 & 3  & 0.10\\
8 & 3 & 10.4 & 8  & 0.24\\
8 & 4 & 17.3 & 15 & 0.30\\ 
\enddata
\end{deluxetable}
\pagebreak

\begin{deluxetable}{ccccccc}
\tabletypesize{\footnotesize}
\tablecaption{Luminosity Results - Global Density\label{table:luminosity}}
\tablewidth{0pt}
\tablehead{
\colhead{Field} 
&\colhead{Luminosity} 
&\colhead{$\Sigma_{\mbox{IPN}}$}
&\colhead{Surface Brightness} 
&\colhead{$\Sigma_{\mbox{galaxies}}$} 
&\colhead{Fraction}
&\colhead{N$_{100}$}\\
&\colhead{($10^{9}$ L$_{\sun}$)}
&\colhead{($10^{5}$ L$_{\sun}$ kpc$^{-2}$)}
&\colhead{($\mu_{\mbox{v}}$ mag~arcsec$^{-2}$)}
&\colhead{($10^{5}$ L$_{\sun}$ kpc$^{-2}$)}
&\colhead{($\Sigma_{\mbox{IPN}}$ / 
$\Sigma_{\mbox{galaxies}}$ + $\Sigma_{\mbox{IPN}}$)}
&\colhead{(galaxies)}}
\startdata
2 & 3.9$^{+1.7}_{-3.3}$ & 8.4 & 27.4 & 15.2 & 0.35 $\pm$ 0.22 & 12\\ 
3h& 9.0$^{+1.2}_{-1.4}$ & 19.9& 26.5 & 107  & 0.16 $\pm$ 0.02 & 26\\ 
3l& 2.0$^{+0.6}_{-0.8}$ & 4.5 & 28.1 & 107  & 0.04 $\pm$ 0.01 & 26\\        
4 & 2.4$^{+1.7}_{-1.7}$ & 8.3 & 27.4 & 13.5 & 0.37 $\pm$ 0.26 & 34\\ 
5 & 3.5$^{+2.0}_{-2.0}$ & 9.5 & 27.3 & 13.5 & 0.41 $\pm$ 0.23 & 11\\ 
6 & 1.7$^{+0.4}_{-0.8}$ & 4.5 & 28.1 & 15.2 & 0.23 $\pm$ 0.08 & 20\\
7 & 6.5$^{+1.1}_{-1.4}$ & 3.2 & 28.4 & 13.5 & 0.19 $\pm$ 0.04 & 31\\
8 & 0 & -- & -- & -- & -- & 21\\
Weighted Avg. & -- & -- & -- & -- & 0.158 $\pm$ 0.03 & -- \\ 
\hline

\enddata
\end{deluxetable}
\pagebreak

\begin{deluxetable}{cccccc}
\tabletypesize{\footnotesize}
\tablecaption{Luminosity Results - Local Density\label{table:localluminosity}}
\tablewidth{0pt}
\tablehead{
\colhead{Field} 
&\colhead{$\Sigma_{\mbox{IPN}}$}
&\colhead{$\Sigma_{\mbox{25}}$} 
&\colhead{Fraction$_{\mbox{25}}$}
&\colhead{$\Sigma_{\mbox{50}}$} 
&\colhead{Fraction$_{\mbox{50}}$}\\
&\colhead{($10^{5}$ L$_{\sun}$ kpc$^{-2}$)}
&\colhead{($10^{5}$ L$_{\sun}$ kpc$^{-2}$)}
&\colhead{($\Sigma_{\mbox{IPN}}$ / 
$\Sigma_{\mbox{25}}$ + $\Sigma_{\mbox{IPN}}$)}
&\colhead{($10^{5}$ L$_{\sun}$ kpc$^{-2}$)}
&\colhead{($\Sigma_{\mbox{IPN}}$ / 
$\Sigma_{\mbox{50}}$ + $\Sigma_{\mbox{IPN}}$)}}
\startdata
2 & 8.4 & 4.2  & 0.67 $\pm$ 0.42 & 18.9 & 0.31 $\pm$ 0.20\\ 
3h& 19.9& 48.8 & 0.29 $\pm$ 0.04 & 23.4 & 0.46 $\pm$ 0.06\\ 
3l& 4.5 & 48.8 & 0.08 $\pm$ 0.02 & 23.4 & 0.16 $\pm$ 0.05\\        
4 & 8.3 & 7.7  & 0.52 $\pm$ 0.36 &  6.0 & 0.58 $\pm$ 0.41\\ 
5 & 9.5 & 11.1 & 0.46 $\pm$ 0.26 &  7.8 & 0.55 $\pm$ 0.31\\ 
6 & 4.5 & 41.4 & 0.10 $\pm$ 0.04 & 13.9 & 0.24 $\pm$ 0.08\\
7 & 3.2 & 15.9 & 0.17 $\pm$ 0.03 & 11.3 & 0.22 $\pm$ 0.04\\
8 & 0 & -- & -- & -- & --\\
Weighted Avg. & -- & -- & 0.133 $\pm$ 0.02 & -- & 0.218 $\pm$ 0.03\\ 
\hline

\enddata
\end{deluxetable}
\pagebreak

\end{document}